\documentclass[a4paper,11pt]{article}
\pdfoutput=1 

\usepackage{jheppub} 

\usepackage[T1]{fontenc} 

\usepackage{amsmath,amssymb,tikz}
\usepackage{graphicx}
 \allowdisplaybreaks
\newcounter{multieqs}

\newcommand{\be}{\begin{equation}}
\newcommand{\ee}{\end{equation}}
\newcommand{\eq}[1]{(\ref{#1})}
\newcommand{\bit}{\begin{itemize}}  \newcommand{\eit}{\end{itemize}}
\newcommand{\ben}{\begin{enumerate}}  \newcommand{\een}{\end{enumerate}}

\newcommand{\bm}[1]{\mbox{\boldmath $#1$}}
\newcommand{\rf}[1]{(\ref{#1})}

\def\bd{\begin{document}}
\def\ed{\end{document}}
 \def\bea{\begin{eqnarray}}
 \def\eea{\end{eqnarray}}
\let\bm=\bibitem

\def\la{\langle}
\def\ra{\rangle}

\def\npb#1#2#3{Nucl. Phys. {\bf{B#1}} #3 (#2)}
\def\plb#1#2#3{Phys. Lett. {\bf{#1B}} #3 (#2)}
\def\prl#1#2#3{Phys. Rev. Lett. {\bf{#1}} #3 (#2)}
\def\prd#1#2#3{Phys. Rev. {D bf{#1}} #3 (#2)}
\def\cmp#1#2#3{Comm. Math. Phys. {\bf{#1}} #3 (#2)}
\def\cqg#1#2#3{Class. Quantum Grav. {\bf{#1}} #3 (#2)}
\def\nppsa#1#2#3{Nucl. Phys. B (Proc. Suppl.) {\bf{#1A}}#3 (#2)}
\def\ap#1#2#3{Ann. of Phys. {\bf{#1}} #3 (#2)}
\def\ijmp#1#2#3{Int. J. Mod. Phys. {\bf{A#1}} #3 (#2)}
\def\rmp#1#2#3{Rev. Mod. Phys. {\bf{#1}} #3 (#2)}
\def\mpla#1#2#3{Mod. Phys. Lett. {\bf A#1} #3 (#2)}
\def\jhep#1#2#3{J. High Energy Phys. {\bf #1} #3 (#2)}
\def\atmp#1#2#3{Adv. Theor. Math. Phys. {\bf #1} #3 (#2)}

\def\N{{\cal N}}
\def\sst{\scriptscriptstyle}
\def\thetabar{\bar\theta}
\def\Tr{{\rm Tr}}
\def\one{\mbox{1 \kern-.59em {\rm l}}}

\def\a{\alpha}      \def\da{{\dot\alpha}}  \def\dA{{\dot A}}
\def\b{\beta}       \def\db{{\dot\beta}}
\def\g{\gamma}  \def\G{\Gamma}  \def\dc{{\dot\gamma}}
\def\d{\delta}  \def\D{\Delta}  \def\ddt{\dot\delta}
\def\e{\epsilon}
\def\ve{\varepsilon}
\def\uve{\upvarepsilon}
\def\f{\phi}    \def\F{\Phi}    \def\vvf{\f}
\def\vphi{\varphi}
\def\h{\eta}
\def\k{\kappa}
\def\l{\lambda} \def\L{\Lambda}
\def\m{\mu} \def\n{\nu}
\def\o{\omega}
\def\p{\pi} \def\P{\Pi}
\def\r{\rho}
\def\s{\sigma}  \def\S{\Sigma}
\def\t{\tau}
\def\th{\theta} \def\Th{\Theta} \def\vth{\vartheta}
\def\X{\Xeta}
\def\z{\zeta}

\def\na{\nabla}

\def\cA{{\mathcal A}} \def\cB{{\cal B}} \def\cC{{\cal C}}
\def\cD{{\cal D}} \def\cE{{\cal E}} \def\cF{{\cal F}}
\def\cG{{\cal G}} \def\cH{{\cal H}} \def\cI{{\cal I}}
\def\cJ{{\mathscr J}} \def\cK{{\cal K}} \def\cL{{\cal L}}
\def\cM{{\cal M}} \def\cN{{\cal N}} \def\cO{{\cal O}}
\def\cP{{\cal P}} \def\cQ{{\cal Q}} \def\cR{{\cal R}}
\def\cS{{\cal S}} \def\cT{{\cal T}} \def\cU{{\cal U}}
\def\cV{{\cal V}} \def\cW{{\cal W}} \def\cX{{\cal X}}
\def\cY{{\cal Y}} \def\cZ{{\cal Z}}

\def\ua{\underline{\alpha}}
\def\uc{\underline{\phantom{\alpha}}\!\!\!\gamma}
\def\um{\underline{\mu}}
\def\ud{\underline\delta}
\def\ue{\underline\epsilon}
\def\una{\underline a}\def\unA{\underline A}
\def\unb{\underline b}\def\unB{\underline B}
\def\unc{\underline c}\def\unC{\underline C}
\def\und{\underline d}\def\unD{\underline D}
\def\une{\underline e}\def\unE{\underline E}
\def\unf{\underline{\phantom{e}}\!\!\!\! f}\def\unF{\underline F}
\def\unm{\underline m}\def\unM{{\underline M}}
\def\unn{\underline n}\def\unN{{\underline N}}
\def\unp{\underline{\phantom{a}}\!\!\! p}\def\unP{\underline P}
\def\unq{\underline{\phantom{a}}\!\!\! q}
\def\unQ{\underline{\phantom{A}}\!\!\!\! Q}
\def\unH{\underline{H}}

\def\As {{A \hspace{-6.4pt} \slash}\;}
\def\bs {{b \hspace{-6.4pt} \slash}\;}
\def\Ds {{D \hspace{-6.4pt} \slash}\;}
\def\Gts {{\Gt \hspace{-6.4pt} \slash}\;}
\def\ds {{\del \hspace{-6.4pt} \slash}\;}
\def\ss {{\s \hspace{-6.4pt} \slash}\;}
\def\ks {{ k \hspace{-6.4pt} \slash}\;}
\def\ps {{p \hspace{-6.4pt} \slash}\;}
\def\xs {{x \hspace{-6.4pt} \slash}\;}
\def\pas {{{p_1} \hspace{-6.4pt} \slash}\;}
\def\pbs {{{p_2} \hspace{-6.4pt} \slash}\;}
\def\cFs {{{\cal F} \hspace{-6.4pt} \slash}\;}
\def\Dss {{D \hspace{-7.5pt} \slash}\;}
\def\dss {{\del \hspace{-7.0pt} \slash}\;}

\def\Ah{{\hat{A}}}
\def\Ch{{\hat{C}}}
\def\Dh{{\hat{D}}}
\def\Gh{{\hat{G}}}
\def\Fh{{\hat{F}}}
\def\Ih{{\hat{I}}}
\def\Jh{{\hat{J}}}
\def\Kh{{\hat{K}}}
\def\Lh{{\hat{L}}}
\def\Ph{{\hat{P}}}
\def\Rh{{\hat{R}}}
\def\Vh{{\hat{V}}}
\def\Xh{{\hat{X}}}

\def\ah{{\hat{\a}}}
\def\bh{{\hat{\b}}}
\def\gh{{\hat{\g}}}
\def\dh{{\hat{\d}}}
\def\rh{{\hat{\r}}}
\def\hh{\hat{h}}
\def\uh{\hat{u}}
\def\xh{\hat{x}}
\def\yh{\hat{y}}
\def\ph{\hat{p}}
\def\xih{\hat{\xi}}
\def\chih{\hat{\chi}}
\def\Psih{\hat{\Psi}}
\def\phih{\hat{\phi}}

\def\psit{\tilde{\psi}}
\def\Psit{\tilde{\Psi}}
\def\Psibt{\tilde{\bar{Psi}}}

\def\st{\tilde{\sigma}}

\def\delt{\tilde{\delta}}
\def\Phit{\tilde{\Phi}}
\def\Phitb{\overline{\tilde{Phi}}}
\def\tht{\tilde{\th}}
\def\lt{\tilde{\l}}
\def\chit{\tilde{\chi}}
\def\phit{\tilde{\phi}}

\def\At{\tilde{A}}
\def\Bt{\tilde{B}}
\def\Ct{\tilde{C}}
\def\Dt{\tilde{D}}
\def\Et{\tilde{E}}
\def\Ft{\tilde{F}}
\def\Gt{\tilde{G}}
\def\Ht{\tilde{H}}
\def\It{\tilde{I}}
\def\Jt{\tilde{J}}
\def\Qt{\tilde{Q}}
\def\Rt{\tilde{R}}
\def\Mt{\tilde{M }}
\def\Nt{\tilde{N}}
\def\St{\tilde{S}}
\def\Vt{\tilde{V}}
\def\Xt{\tilde{X}}
\def\at{\tilde{a}}
\def\ct{\tilde{c}}
\def\dt{\tilde{d}}
\def\htt{\tilde{h}}
\def\ft{\tilde{f}}
\def\gt{\tilde{g}}
\def\pt{\tilde{p}}
\def\qt{\tilde{q}}
\def\vt{\tilde{v}}
\def\nt{\tilde{n}}
\def\ut{\tilde{u}}
\def\wt{\tilde{w}}
\def\zt{\tilde{z}}
\def\xt{\tilde{x}}
\def\yt{\tilde{y}}
\def\Psit{\tilde{\Psi}}
\def\vphit{\tilde{\varphi}}
\def\tD{\tilde{\D}}

\def\eb{\bar{\epsilon}}
\def\delb{\bar{\partial}}
\def\thb{\bar{\theta}}
\def\mub{\bar{\mu}}
\def\lamb{\bar{\l}}
\def\psib{\bar{\psi}}
\def\sb{\bar{\sigma}}
\def\xib{\bar{\xi}}
\def\chib{\bar{\chi}}

\def\Psib{\bar{\Psi}}
\def\Phib{\bar{\Phi}}
\def\Lamb{\bar{\Lambda}}
\def\Sb{{\overline \Sigma}}
\def\cb{\bar{c}}
\def\hb{\bar{h}}
\def\qb{\bar{q}}
\def\wb{\bar{w}}
\def\ub{\bar{u}}
\def\zb{{\bar{z}}}
\def\Hb{\bar{H}}
\def\Qb{{\bar Q}}
\def\Omegab{\overline{\Omega}}
\def\ob{\overline{\omega}}

\def\Ab{{\overline A}} \def\Bb{{\overline B}} \def\Cb{{\overline C}}
\def\Db{{\overline D}} \def\Eb{{\overline E}} \def\Fb{{\overline F}}
\def\Gb{{\overline G}}
\def\Ib{{\overline I}}
\def\Jb{{\overline J}} \def\Kb{{\overline K}} \def\Lb{{\overline L}}
\def\Mb{{\overline M}} \def\Nb{{\overline N}} \def\Ob{{\overline O}}
\def\Pb{{\overline P}}  \def\Rb{{\overline R}}
 \def\Tb{{\overline T}} \def\Ub{{\overline U}}
\def\Vb{{\overline V}} \def\Wb{{\overline W}} \def\Xb{{\overline X}}
\def\Yb{{\overline Y}} \def\Zb{{\overline Z}}

\def\fb{{\overline f}}
\def\gb{{\overline g}}
\def\mb{{\overline m}}
\def\lb{{\overline l}}
\def\yb{{\overline y}}

\def\ba{{\bf a}}
\def\bk{{\bf k}}
\def\bl{{\bf l}}
\def\bp{{\bf p}}
\def\bq{{\bf q}}
\def\br{{\bf r}}
\def\bt{{\bf t}}
\def\bu{{\bf u}}
\def\bv{{\bf v}}
\def\bx{{\bf x}}
\def\by{{\bf y}}
\def\bA{{\bf A}}
\def\bB{{\bf B}}
\def\bR{{\bf R}}
\def\bV{{\bf V}}

\def\bz{{\boldsymbol{\zeta}}}

\def\bone{{\bf 1}}

\def\va{{\vec a}}
\def\vk{{\vec k}}
\def\vp{{\vec p}}
\def\vq{{\vec q}}
\def\vx{{\vec x}}
\def\vy{{\vec y}}
\def\vu{{\vec u}}
\def\vv{{\vec v}}
\def \vH{{\vec H}}
\def \vg{{\vec g}}

\def\vs{{\vec \sigma}}
\def\vtau{{\vec \tau}}

\def\frA{\mathfrak{A}}
\def\frB{\mathfrak{B}}
\def\frC{\mathfrak{C}}
\def\frD{\mathfrak{D}}
\def\frE{\mathfrak{E}}
\def\frF{\mathfrak{F}}
\def\frG{\mathfrak{G}}
\def\frH{\mathfrak{H}}
\def\frM{\mathfrak{M}}
\def\frN{\mathfrak{N}}
\def\frR{\mathfrak{R}}
\def\frW{\mathfrak{W}}

\def\fra{\mathfrak{a}}
\def\frb{\mathfrak{b}}
\def\frf{\mathfrak{f}}
\def\frg{\mathfrak{g}}
\def\frh{\mathfrak{h}}
\def\frl{\mathfrak{l}}
\def\frs{\mathfrak{s}}
\def\fri{\mathfrak{i}}
\def\frj{\mathfrak{j}}

\def\ma{\mathfrak{a}}
\def\mg{\mathfrak{g}}
\def\mh{\mathfrak{h}}
\def\mR{\mathfrak{R}}
\def\mN{\mathfrak{N}}

\newcommand{\nn}{{\nonumber}}

\def\d{\delta}\def\D{\Delta}\def\ddt{\dot\delta}

\def\pa{\partial} \def\del{\partial}
\def\xx{\times}
\def\uno{\mbox{1 \kern-.59em {\rm l}}}

\def\trp{^{\top}}
\def\inv{^{-1}}
\def\dag{\dagger}
\def\pr{^{\prime}}

\def\rar{\rightarrow}
\def\lar{\leftarrow}
\def\lrar{\leftrightarrow}

\newcommand{\0}{\,\!}      
\def\one{1\!\!1\,\,}
\def\im{\imath}
\def\jm{\jmath}

\newcommand{\tr}{\mbox{tr}}
\newcommand{\slsh}[1]{/ \!\!\!\! #1}

\def\vac{|0\rangle}
\def\lvac{\langle 0|}

\def\hlf{\frac{1}{2}}
\def\ove#1{\frac{1}{#1}}
\newcommand{\hot}[1]{\frac{#1}{2}}

\def\Box{\square}
\def\CC {\mathbb{C}}
\def\FF {\mathbb{F}}
\def\RR{\mathbb{R}}
\def\NN{\mathbb{N}}
\def\ZZ{\mathbb{Z}}
\def\bb#1{{\bf #1}}
\def\bcomment#1{}
\def\bfhat#1{{\bf \hat{#1}}}
\def\VEV#1{\left\langle #1\right\rangle}

\newcommand{\ex}[1]{{\rm e}^{#1}} \def\ii{{\rm i}}

\newcommand{\lrbrk}[1]{\left(#1\right)}
\newcommand{\lrsbrk}[1]{\left[#1\right]}
\newcommand{\sfrac}[2]{{\textstyle\frac{#1}{#2}}}

\def\stw{{\sqrt{2}}}

\def\rf {{\rm f}}
\def\ri {{\rm i}}
\def\rj {{\rm j}}
\def\rn {{\rm n}}
\def\rk {{\rm k}}
\def\rl {{\rm l}}
\def\rr {{\rm r}}
\def\rs {{\scriptscriptstyle \rm S}}
\def\rt {{\scriptscriptstyle \rm T}}

\def\rQ {{\scriptscriptstyle \rm \cQ}}
\def\rR {{\scriptscriptstyle \rm \cR}}

\def\cQb{{\cal \Qb}}
\def\cRb{{\cal \Rb}}
\def\cWb{{\cal \Wb}}

\def\fd {{\rm N}}
\def\afd {{\overline{\rm N}}}

\def \II {I\hspace{-.1em}I\hspace{.1em}}
\def \IIA {\mbox{\II A\hspace{.2em}}}
\def \IIB {\mbox{\II B\hspace{.2em}}}
\def \gs {g^s}
\def \ls {\lambda^s}

\def \I {{\cal I}}
\def \qs {q\hspace{-.53em}/\hspace{.15em}}
\def \ks {k\hspace{-.53em}/\hspace{.15em}}
\def \YM {{\mbox{\tiny YM}}}
\def \gym {g_{\YM}}

\def \Lc {\L_c}
\def\IR{\relax{\rm I\kern-.18em R}}
\def \id {{\bf 1}}

\def\cci{\ell}
\def\ccj{\ell'}

\def\bbq{\pmb{q}}

\newcommand{\aei}{\it Max Planck Institute for Gravitational Physics
(Albert Einstein Institute)\\ Am M\"uhlenberg 1, 14476 Golm,
Germany}

\newcommand{\nthu}{{\it Department of Physics, National Tsing-Hua
  University,
Hsinchu 30013, Taiwan}}

\newcommand{\ncts}{{\it Physics Division, National Center for Theoretical Sciences,
National Tsing-Hua University, Hsinchu 30013, Taiwan}}

\newcommand{\ictsustc}{{\it Interdisciplinary Center for Theoretical Study,
University of Science and Technology of China,\\
Hefei, Anhui 230026, People's Republic of China}}

\newcommand{\sysu}{{\it School of Physics and Astronomy, Sun Yat-Sen University, Zhuhai, 519082, China}}

\begin{document}

\title{\boldmath  Boundary  String Current \& Weyl Anomaly \\in 
  Six-dimensional Conformal Field Theory}



\author[a,b]{Chong-Sun Chu,}
\author[c]{Rong-Xin Miao }


\affiliation[a]{Physics Division, National Center for Theoretical Sciences,\\
National Tsing-Hua University, Hsinchu 30013, Taiwan}
\affiliation[b]{ Department of Physics, National Tsing-Hua
  University,
Hsinchu 30013, Taiwan}
\affiliation[c]{ School of Physics and Astronomy, Sun Yat-Sen University, Zhuhai, 519082, China}

\emailAdd{cschu@phys.nthu.edu.tw}
\emailAdd{miaorx@mail.sysu.edu.cn}

\preprint{NCTS-TH/1818}

\abstract{  It was recently discovered that for a boundary system
  in the presence of a background
  magnetic field, the quantum fluctuation of the vacuum would
  create a non-uniform magnetization density for the vacuum and a magnetization
  current is induced in the vacuum \cite{Chu:2018ksb}.
  It was also shown that this ``magnetic
  Casimir effect'' of the vacuum is closely related to another quantum
  effect of the
  vacuum,  the Weyl anomaly. Furthermore, the phenomena can be
  understood in terms of the holography of the
  boundary system \cite{Chu:2018ntx}.
  In this paper, we generalize this four dimensional
  effect to six dimensions. We use the AdS/BCFT holography to show  
  that  in the presence of a 3-form magnetic
  field strength $H$, 
  a string current is induced in a six dimensional boundary conformal field
  theory. This  allows us to determine the gauge field contribution to the
  Weyl anomaly in six dimensional  conformal field theory in a
  $H$-flux background.
  For the (2,0) superconformal field theory of $N$ M5-branes, the
  current has a magnitude proportional to $N^3$ for large $N$.
  This suggests that
  the degree of freedoms scales as $N^3$ in the (2,0) superconformal
  theory of $N$ multiple M5-branes. The prediction we have for the Weyl anomaly
  is a new criteria that the (2,0) theory
  should satisfy. }

\maketitle
\flushbottom

\section{Introduction}

The decoupling limit of $N$ coincident M5-branes 
is given by an interacting (2,0) superconformal
theory in 6 dimensions \cite{m5-02}.
For a single M5-brane, the low energy theory
is known and is given by a free theory of tensor multiplet
\cite{howe,howe1,PS,schw1,pst,pst1}. The multiple
M5-branes theory is much more complicated and it is not expected to have a
fundamental in terms of a local  Lagrangian description
\footnote{However just like in supergravity or hydrodynamics,
  it is perfectly sensible to look for a classical Lagrangian
  description for the effective dynamics of the (2,0) fields in the
  long wavelength limit. In this context, see for example \cite{CK}
where
a set of non-Abelian self-dual equation has been 
constructed and proposed as the classical equation of motion for the self-dual
tensor gauge field in the 
low energy effective theory of the multiple M5-branes theory, with
various supporting evidences obtained in
\cite{CK,CKV,CV,CI,CI11,CI22}.
We remark  however that supersymmetry is  missing in  this construction.
}.
There exists a number of  proposals for the fundamental formulation of the
six dimensional (2,0)
theory: most notably, these include the Discrete Light-Cone Quantisation
definition
based on quantum mechanics on the moduli space of instantons \cite{qm1,qm2},
a definition based on deconstruction from four dimensional superconformal,
quiver field theories \cite{dec}, and the conjecture that the (2,0) theory
compactified on a circle is equivalent to the five dimensional maximally
supersymmetric Yang-Mills theory \cite{dou,lam2}.
And despite an extensive amount
of work on this topic, see for example,
\cite{Chu:2009iv,Chu:2009ms,Chu:2011fd,Ho:2014eoa, Ho:2011ni,
Samtleben:2011fj,Kim:2011mv,Kallen:2012zn,Bern:2012di,
Cordova:2015vwa,Beem:2015aoa,Hosomichi:2012ek,Kallen:2012va,
Kim:2012ava,Kim:2012gu,Kim:2012qf,Saemann:2012uq,Saemann:2016sis},
the field theoretic description of the
multiple M5-branes system remains mysterious.
In addition to consistency and symmetry
requirement, the fundamental theory, no matter how
it is defined, should reproduce properties that are expected of
the multiple M5-branes system. For example, it should describe a
non-trivial interacting theory of (2,0) superconformal multiplets. It should
contain BPS states of self-dual strings which corresponds to boundaries
of M2-branes ending on the stack of M5-branes
\cite{Strominger:1995ac,Dijkgraaf:1996hk}.
It should explain the $S$ duality of the
$\cN=4$ supersymmetric Yang-Mills theory \cite{Tachikawa:2011ch}.
It should also make apparent the $N^3$
entropy behaviour \cite{Klebanov:1996un}. In particular it should explain
whether this is due to novel degrees of freedom of the (2,0) theory or not.
One of the motivations of this work has been
to add new criteria to the list by
uncovering new physical properties of the multiple
M5-branes system.

To this end, we recall an useful approach that has been known
to work very well in the past is to
introduce a boundary to the system. For example, the form of the
noncommutative geometry, including the relation between closed string
and open string metric, on D-branes worldvolume can be derived
by considering open strings  ending on the D-branes \cite{Chu:1998qz,Chu:1999gi}.
Therefore we are motivated to consider a M5-branes system with boundary. The
resulting low energy theory is a boundary conformal field theory (BCFT),
which is
a very interesting class of theories by itself.

Boundary conformal field theory 
\cite{Cardy:2004hm,McAvity:1993ue} 
describes the fixed point of renormalization group (RG) flow in boundary
quantum field theory and has important
applications in  quantum field theory, string theory
and condensed matter system such as,
for example, renormalization
group flows and critical phenomena \cite{Cardy:2004hm} or
the topological
insulator \cite{Hasan:2010xy}. 
For general shape of the boundary, traditional perturbative analysis of BCFT
becomes exceedingly complicated. 
In addition to traditional
field theory techniques, see, e.g.
\cite{
  Fursaev:2015wpa,Herzog:2015ioa,Miao:2017aba,Herzog:2017kkj,
Jensen:2017eof, 
Kurkov:2017cdz,Kurkov:2018pjw,Rodriguez-Gomez:2017kxf},
the need of a
non-perturbative approach using symmetries 
or dualities is evident. 
A non-perturbative holographic dual description to BCFT was initiated by Takayanagi
in \cite{Takayanagi:2011zk} and later developed for general shape of boundary
geometry in \cite{Miao:2017gyt,Chu:2017aab}. The duality has been extensively
studied in the literature, with many interesting results obtained. See, for example
\cite{Fujita:2011fp,Nozaki:2012qd,Astaneh:2017ghi,
  Flory:2017ftd, Bhowmick:2017egz, Herzog:2017xha,
  Seminara:2017hhh,Chang:2018pnb,Seminara:2018pmr,Miao:2018qkc,Andrei:2018die}.

The Casimir effect is one of the most well known manifestation of the
quantum fluctuation of 
vacuum in the presence of boundary
\cite{Casimir:1948dh,Plunien:1986ca,Bordag:2001qi}.
Recently the Casimir effects has been
analyzed in full generality for arbitrary shape of boundary and
for arbitrary spacetime metric.
Universal relations between the Casimir coefficients
which determine the  near boundary behaviour
of the renormalized stress tensor 
and the boundary central charge in a boundary conformal
field theory have been discovered \cite{Miao:2017aba}.
The analysis has also been extended to $U(1)$ current in BCFT
\cite{Chu:2018ksb}. It was
found that when an external magnetic field
is applied, the vacuum of BCFT will get magnetized  and
a magnetization current get induced in the
vicinity of the boundary. In analogous to the standard Casimir effect
which is a manifestation of the mechanical property of the vacuum, this
effect  is a
manifestation of the magnetic property of the quantum vacuum and
may be refereed to as a 
{\it magnetic Casimir effect}. The generalization of
this effect to higher dimensions
was another motivation of this work.

The above described effects for the stress
tensor and the $U(1)$ current can be derived from the AdS/BCFT holography
\cite{Miao:2017aba,Chu:2018ntx}.
From the field theory point of view, they can also be derived from
the Weyl anomaly of the BCFT \cite{Miao:2017aba,Chu:2018ksb}.
 Consider a CFT with partition function $Z[g_{\m\n}]$ and the effective action
$W[g_{\m\n}] = \ln Z[g_{\m\n}]$. 
The scaling symmetry of CFT is generally broken due to quantum effects
and the breaking is measured by the Weyl anomaly
\be
\mathcal{A} := \del_\vphi W[e^{2 \vphi} g_{\m\n}] \big|_{\vphi=0} = \int_M \la
T_\m^\m \ra.
\ee
The metric contribution to the Weyl anomaly is well understood. For example in
even dimensions, the bulk part of the Weyl anomaly takes the form
\cite{Deser:1993yx}
\be
\la T_\m^\m \ra = \frac{1}{(4\pi)^{d/2}} \left(
\sum_j c_{dj}I^{(d)}_j - (-1)^{\frac{d}{2}}a_d E_d
\right).
\ee
Here 
$E_d$ is the Euler density in $d$ dimensions,
$I^{(d)}_j$ are independent Weyl invariants of weight $-d$ and the subscript $j$
labels the Weyl invariants. The boundary terms of the Weyl anomaly has also been
studied and classified recently in \cite{Herzog:2015ioa}. In general, in addition
to a nontrivial background metric, one may also turn on a gauge field background
and the loops of matter fields  will
give  a Weyl anomaly. For example in 4 dimensions,
vector gauge field (Abelian or non-Abelian)
is classically conformal and there is a  Weyl anomaly \cite{Peskin}
\be \label{anom-4d}
\la
T_\m^\m \ra = b\; \tr F^2.
\ee
Here $b = \beta(g)/2g^3$ and $\beta(g)$ is the beta function of the theory
$S= -1/(4g^2) \int \tr F^2$.
For higher $d=2n$ even dimensions,
a $n$-form gauge field $H$ is conformal invariant classically. One can
expect a background of $H$-flux will give rises to a Weyl anomaly. 
However since
we do not even know what the higher rank gauge field couple to and how, 
let alone the quantisation, 
nothing is known about the possible form of this anomaly.
The wish to say something about this Weyl anomaly
from background of higher form gauge
field has been another major motivation of this work.

That this goal can be achieved follows from the observation
in \cite{Chu:2018ksb} that
the gauge part of the
Weyl anomaly is intimately related with the induced current,
see \eq{key} for 4-dimensions. Similar
anomaly-current relation can be straightforwardly established for
higher dimensions. 
In the case of six dimensions where we are interested
in, we can introduce a boundary to the CFT and first use
AdS/BCFT to compute the induced string current,
and then use this result and the anomaly-current relation to
determine the gauge field contribution to the
Weyl anomaly in  6-dimensional CFT.

The plan of this paper is as follows. In section 2, we first
review the phenomena
of induced current in 4 dimensions. We then generalize it
to six dimensions. We show that the use of
symmetries and conservation law of the theory allows us to fix,
up to a few numerical coefficients,
the form of the
one point function of a conserved current in the presence of a background
of  3-form
flux. In section 3, we use AdS/BCFT holography to determine the form of the
induced current. The result is consistent with
the form obtained by the field theory analysis.
In section 4, we generalize the relation between Weyl anomaly
and the conserved current to six dimensional BCFT. Using this relation
and the result of the induced
current from AdS/BCFT  as input, we obtain the contribution of
the 3-form field strength  to the Weyl anomaly in six
dimensional CFT.
For the system of maximal (2,0) supersymmetric multiple $N$ M5-branes,
the current and the Weyl anomaly are found to be
proportional to $N^3$
\footnote{The $N^3$ dependence has also been found for the
gravitational contribution to the  conformal
  anomaly in the Coulomb 
branch of the (2, 0) theory by relating the Coulomb branch interactions in six 
dimensions to interactions in four dimensions using supersymmetry
\cite{Ganor:1997jx,Maxfield:2012aw,Beem:2014kka,Cordova:2015vwa}.}. 
This provides some evidence that the fundamental degree of
freedoms of the (2,0) theory of $N$ M5-branes scales like $N^3$.

\section{Boundary String Current}

Consider a boundary conformal field theory (BCFT) defined on a 
manifold $M$ with boundary $P$.  
In \cite{Chu:2018ksb}, we have shown that for 4-dimensions, the vacuum
expectation value of the
renormalized
current
$J_\m$ has the asymptotic expansion near the boundary at $x=0$,
\be \label{J4d}
\la J_\m \ra = \frac{\a_1}{x}  F_{\m \l} n^\l + \cdots,
\ee
when a background gauge field strength $F_{\m\n}$ is turned on.
Here $\cdots$ denotes
terms that are less singular.
It was 
shown that the current \eq{J4d} 
is related to the Weyl
anomaly \eq{anom-4d} and the coefficient $\a_1$ is completely
determined in terms of the beta function of the theory. It was also
understood
that the current \eq{J4d} is a consequence of the
magnetization of the vacuum
which arises from the quantum fluctuation of the vacuum in the
presence of the boundary.
Here we are interested in  generalizing this phenomena of near boundary current
to higher dimensional BCFT in the presence of a  higher form gauge field
background.

Let us consider a 6-dimensional BCFT with gauge symmetry
defined on a manifold $M$.
Yang-Mills gauge
field is not conformal invariant in six dimensions,
instead a 2-form gauge field $B_{\m\n}$ is. 
For simplicity, we consider Abelian gauge field here. 
The 2-form gauge potential is naturally coupled to the worldsheet
$\S$ of a string
with the  minimal coupling
\be \label{IB}
I_B = \int_\S B = \int_M J^{\m\n} B_{\m\n}  
\ee
where 
\be
J^{\m\n} = \l \e^{\a\b} \frac{\del X^\m}{\del \s^\a} \frac{\del X^\n}{\del \s^\b}
\d^{(4)} (X - X(\s^a))
\ee
is a two-form string current that arises from the
motion of the string and
$\l$ is the string charge density.
Next let us introduce a boundary $P = \del M$. This breaks the bulk conformal
symmetry and the one point function of the current can become nontrivial now. 
As the current
$J_{\m\n}$ has a mass dimension 4,  the vacuum expectation value of
the renormalized current generally takes the form
\be \label{J-asym}
\la J_{\m\n} \ra
= \frac{1}{x} J_{\m\n}^{(1)} + \log x J_{\m\n}^{(0)} + \cdots 
\ee
near the boundary.
Here we have used  gauge invariance and the conservation law 
\be \label{DJ}
D_\m J^{\m\n} =0
\ee
to rule out  terms like
$ J_{\m\n}^{(4)}/x^4,  J_{\m\n}^{(3)}/x^3,  J_{\m\n}^{(2)}/x^2$. In \eq{J-asym}, 
$\cdots$ denotes terms that are regular at $x=0$, and 
$J_{\m\n}^{(1)}$ and $J_{\m\n}^{(0)}$ are functions of
dimension 3 and 4 respectively.
Their form are constrained by
\eq{DJ} and the Lorentz and gauge symmetries of the theory.
For example, one
can easily determine that
\be \label{J1}
J_{\m\n}^{(1)} = \a_1 H_{\m\n \l} n^\l + \a_2 \cD_{[ \m} \cD_{\n ]} k +
\a_3 \cD_{[ \m} \cD_\l k^\l{}_{\n ]} +\a_4 \cD_\l \cD_{[ \m} k^\l{}_{\n ]},
\ee
where $H_{\m\n \l}, n_\m, \cD_\m, k_{\m\n}$ are respectively the background
3-form field
strength, normal vector to the boundary, induced covariant derivative and the
extrinsic curvature of the boundary.
The coefficients $\a_i$ are arbitrary and contain important physical
information of the theory. In \cite{Chu:2018ksb} it was shown that, for
four dimensions, the near boundary 
asymptotic
form of the standard current $J_\m$ is completely determined by the
background field strength of the Weyl anomaly. It was also shown in
\cite{Chu:2018ntx}
that the near boundary current can also be determined using
the AdS/BCFT holography. For six dimensions, the
background gauge field part of the Weyl anomaly is unknown. Therefore let us
proceed first with the holographic analysis and determine the
near boundary current
using boundary holography.

\section{Holographic Boundary Current}

\begin{figure}[t]
\centering
\includegraphics[width=8cm]{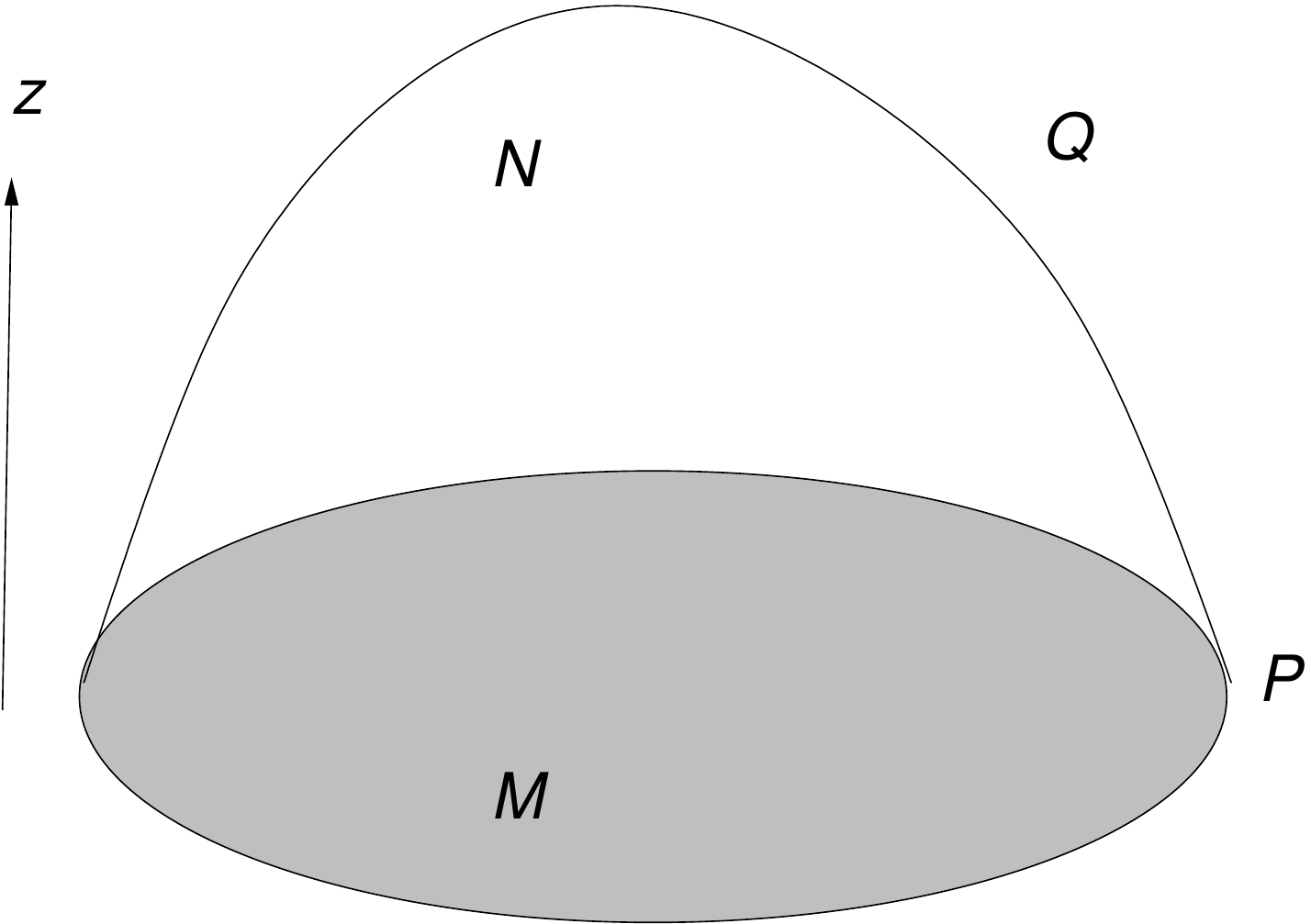}
\caption{BCFT on $M$ and its dual $N$}
\label{MNPQ}
\end{figure}

Holographic dual of BCFT was originally introduced by
Takayanagi
\cite{Takayanagi:2011zk}. The idea is to extend the $d$
dimensional manifold $M$ to a $(d+1)$-dimensional asymptotically AdS
space $N$ such  that $\partial N= M\cup Q$, where $Q$ is a $d$
dimensional manifold with boundary $\partial Q=\partial
M=P$. 
According to the
standard AdS/CFT arguments,
the asymptotic boundary behaviour of a bulk field $\phi$ in $AdS$ generates
the expectation value of local operator $\cQ$ in the CFT. In our case,
let us consider a 2-form tensor gauge field in the bulk
described by the gravitational action
\be \label{action}
I =
\frac{1}{16 \pi G_N}
\int_N \left(R-2 \Lambda - \frac{1}{6q^2} \cH_{LMN}^2\right).
  \ee
  Here $G_N$ is the Newton constant in 7 dimensions,
$1/q^2$ is a dimensional constant of length dimension 4,  
 and $\cH = d \cB$. $\cB$ is the bulk gauge field whose boundary value is
 given by the gauge field $B$ on the boundary $M$.
 Note that $B$ is completely
 arbitrary and does not need to satisfy any equation of motion.
 The bulk indices are denoted by the capital Roman letters
 $L,M,N = 0, 1, \cdots, 6$
 and the indices of the 6-dimensional manifolds $M$ and $Q$ are
 denoted by Greek letters $\mu, \nu$ etc. It should be clear from the context
 whether we are referring to the manifold $M$ or $Q$.
 The existence of the tensor gauge field in the bulk dictate the presence of a
 2-form current $J_{\m\n}$ in the CFT whose expectation value is determined by
the generating relation,
\be
Z_{\rm string} [B_{\m\n}] = \left\la e^{\int_M J^{\m\n} B_{\m\n}} \right\ra.
\ee
In field theory, the current is constructed as the Noether current of some
global Abelian symmetry.
In this paper, we will be interested in the SUGRA limit where
the string partition function is given by the SUGRA action \eq{action}.

 The new ingredient in AdS/BCFT is that the SUGRA action isn't defined until
the shape $Q$ is known.
According to Takayanagai \cite{Takayanagi:2011zk}, $Q$ is
determined as an extremal configuration of the supergravity action 
with respect to change of $Q$:
\be\label{action1}
  16 \pi G_N I = \int_N \left(R-2 \Lambda  - \frac{1}{6q^2} \cH_{LMN}^2\right)
  +  2\int_M K + 2\int_Q (K-T) +2\int_P \theta.
\ee
Here the constant
parameter $T$ is a measure of the boundary degree of freedom of the
BCFT. 

To specific the variational principle, one needs to fix the boundary condition
for $Q$.
As $N$ is of codimension one, the location of $Q$ is determined by a single
function. A 
  consistent model of holographic BCFT
  was found by  considering a mixed BC on $Q$ and the
  following  trace condition \cite{Miao:2017gyt,Chu:2017aab}
 \be
K = \frac{d}{d-1} T \label{NBC1-g}
\ee
was obtained.
The employment of a mixed BC  is a reasonable assumption
  if one think of $Q$ as a brane and then there should be a single embedding
  equation for it
  \footnote{We thank Juan Maldacena for suggesting this interpretation.}.
  In addition, we impose a Neumann boundary condition for the gauge field,
  \be
\mathcal{H}_{LMN}\; n_Q^{L}\Pi^{M}_{\ \a}\Pi^{N}_{\ \b}=0.
\label{NBC-B}
\ee
Here $n_Q$ is the inward-pointing normal vector on $Q$,
the beginning Greek letters $\a,\b$ etc denote indices on $Q$, and $\Pi$ is
the projection operator which gives the vector field and metric on $Q$:
$\bar{B}_{\a\b}=\Pi^{M}_{\ \a}\Pi^{N}_{\ \b}\mathcal{B}_{MN}$ and
$\gamma_{\alpha\beta}=\Pi^{M}_{\ \alpha}\Pi^{N}_{\ \beta}G_{MN}$.  
  We note that \cite{Miao:2017aba} the manifold $N$ is actually singular
  since 
  the normal of $N$ is discontinuous at the junction $P$. Due to
  this discontinuity,
  an expansion in small $z$ in the form of
Fefferman-Graham (FG) asymptotic expansion \cite{FG}
  would not be  sufficient, and 
  one needs to have a full analytic control of the metric near $P$,
  i.e. near $z=0=x$. The need of a non-FG expanded bulk metric
was already anticipated in \cite{Nozaki:2012qd}.
The  general form of this non-FG expanded bulk metric
  that is analytic near $P$ 
  was  successfully constructed in \cite{Miao:2017aba} by
  considering
  an expansion in  small exterior curvature of the boundary surface $P$.
  Moreover it was found 
 that  by using the non-FG
 expansion of the metric in the bulk, the tensor embedding equation
\be
K_{\alpha\beta}-(K-T)\gamma_{\alpha\beta} =0 \label{NBC-g}
\ee
for $Q$ as proposed originally by Takayanagai \cite{Takayanagi:2011zk}
is also  consistent  \cite{Miao:2017aba}: with the tensor model \eq{NBC-g}
considered as a special case of the scalar model \eq{NBC1-g}.

Now back to our system and
let us solve for the shape  $Q$ and the gauge field configuration.
Let us denote the 7-dimensional
bulk indices by
$S =(z,\mu)$, and the 6-dimensional field theory indices by
$\mu =(x,a)$ with
$a= 0,1, \cdots, 4$.
For simplicity, let us consider the case of a  flat half space $x\geq 0$.
The bulk metric reads
\begin{equation}\label{AdSmetric}
ds^2=R^2 \frac{dz^2+dx^2+\delta_{ab}dy^ady^b}{z^2},
\end{equation}
where $R$ is the $AdS_7$ radius.
In this case, \eq{NBC-g} reduces to \eq{NBC1-g}, and $Q$ is given by
\cite{Takayanagi:2011zk}
\begin{equation}\label{Q}
\ x=- z \, \sinh\frac{\rho}{R},
\end{equation}
where we have re-parametrized $T=\frac{5}{R} \tanh \frac{\rho}{R}$.

As for
the solution for the gauge field, we will consider the situation of having
a constant field strength in the BCFT.
Due to the
planar symmetry of the boundary, we consider
$\cB_{MN}$ that depends
only on the coordinates $z$ and $x$. The field equations
$\nabla_{L}\mathcal{H}^{LMN}=0$ can be solved with
non-vanishing components
$\mathcal{B}_{za} =\mathcal{B}_{za}(z)$,
$\mathcal{B}_{xa} =\mathcal{B}_{xa}(x)$,
$\cB_{ab} = \cB_{ab}(z,x)$,
and with
$\cB_{ab}$ satisfying,
\begin{eqnarray}
  \label{EOM-B}
z \del_z^2  \cB_{ab} + z \del_x^2 \mathcal{B}_{ab} -\del_z \cB_{ab} =0.
\end{eqnarray}
To solve this, let us take the ansatz
\be \label{cB-ansatz}
\cB_{ab} = \sum_{n=0}
x^n f_n(\frac{z}{x}) B_{ab}^{(n)},
\ee
where $f_n(0) =1$ 
so that $\cB_{ab}$ reduces to the gauge field $B_{ab}$
at the AdS boundary $z=0$. Here the constants $B_{ab}^{(n)}$'s are the
expansion coefficients of $B_{ab}$ about the boundary $x=0$:
\be
B_{ab} = \sum_{n=0}
x^n B_{ab}^{(n)}.
\ee
Considering the case of a constant field strength $H_{xab}$ in BCFT.
In this case, we have a non-vanishing $B_{ab}^{(1)}$ given by
\be
B^{(1)}_{ab} = H_{xab} 
\ee
and the equation of motion \eq{EOM-B} has the solution
$\cB_{ab} = x f(\frac{z}{x}) H_{xab}$
with $f(s) = (1-c_1) + c_1 \sqrt{1+s^2}$.
The boundary condition \eq{NBC-B} then imposes that $c_1=1$ and
\be
\cB_{ab} = H_{xab} \sqrt{x^2+z^2}.
\ee
This give rises to the non-vanishing components
\be
\cH_{zab} = H_{xab} \frac{z}{ \sqrt{x^2+z^2}},\qquad
\cH_{xab} =  H_{xab} \frac{x}{ \sqrt{x^2+z^2}}
\ee
for the bulk field strength.
From the gravitational action \eq{action}, one then derive the
holographic current
\be \label{J-B}
\la J^{ab} \ra = \lim_{z\to 0} \frac{\d I }{\d \cB_{ab}}
= b  \frac{ H_{xab}}{x},
\ee
where  
\be \label{b-hol}
b = - \frac{R}{16\pi G_N q^2}
\ee
is a constant.
It is remarkable that the current \eq{J-B}
is independent of the parameter $T$, showing that
the boundary current in 6d BCFT is independent of boundary conditions.

\section{Weyl Anomaly from Boundary Current}

Recall that for 4-dimensional BCFT, the following 
relation
can be established  \cite{Chu:2018ksb}
\be \label{key}
(\d \mathcal{A} )_{\del M} = \Big(\int_{M_\e}
J^\m \d A_\m
\Big)_{\log \frac{1}{\e}}.
\ee
Here $\e>0$ is a UV regulator
and $M_\e$ is the regulated manifold with $x\geq \e$. The relation \eq{key}
relates the boundary term of the variation of the Weyl anomaly
under an arbitrary variation
of the vector gauge field $\d A_\m$
with the coefficient of the
logarithmic divergent term of the regulated integral on the right
hand side. In general the Weyl anomaly can be computed from the quantum effects
of matter loops on the path integral with external gauge fields.
The relation \eq{key} then allows one to determine the form
of the current \cite{Chu:2018ksb} (and also the stress tensor
\cite{Miao:2017aba})
near the boundary. Vice
versa, one may also use the current as determined by holography as input and
use it to determine the Weyl anomaly.
The results are of course all consistent with each
other.

In higher dimensions, the gauge field contribution to the Weyl anomaly is
unknown.
Nevertheless,
even without any knowledge of the path integral or how the higher rank gauge
field,
one can easily generalize the analysis of \cite{Chu:2018ksb}
and establish
a similar relation \eq{key} between the Weyl anomaly and the boundary current.
couples to the other fields of the system. 
To
be concrete, let us consider $d=6$. In this case,
for a conserved 2-form current $J^{\m\n}$, $\del_\m J^{\mu\nu}=0$ 
coupled to an external 2-form tensor
gauge field $B_{\m\n}$ with the coupling \eq{IB},
we have the relation
\be \label{key1}
(\d \mathcal{A} )_{\del M_\e} = \Big(\int_{M_\e}
J^{\m\n} \d B_{\m\n}
\Big)_{\log \frac{1}{\e}}.
\ee
For completeness, we give a proof of \eq{key1} in the appendix of the paper.
Using the holographic result \eq{J-B}, and
for a generic field strength,
one can verify that \eq{key} is satisfied
with $\mathcal{A}$ given by
\be \label{anom-6d}
\mathcal{A}=\int_M \;
\frac{b}{6} H_{\m\n\l}^2
\ee
We remark
that one can also use the AdS/BCFT to compute the holographic stress tensor
and the Weyl anomaly \cite{Miao:2017gyt,Chu:2017aab}.
The same result is obtained.
This is our prediction for the form
of the gauge field contribution in the Weyl anomaly in 6d CFT with tensor gauge
field.

An interesting application is in the theory of multiple M5-branes in
M-theory. Consider a system of $N$ coincident M5-branes in flat space.
Although the field theoretic description of the six-dimensional
(2,0) superconformal
field theory is unknown,
nevertheless it is possible to give a  holographic
description of the system
by M-theory on an $AdS_7 \times S^4$ background. The supergravity background is
given by
a constant 4-form field strength and the metric
\be
ds^2 = R^2 \frac{dz^2 + dx_6^2}{z^2} + R'{}^2 d\Omega_4^2
\ee
where  $R = 2 (\pi N)^{1/3} l_{11}$ is the AdS radius, $R' =R/2$ and
$l_{11}$ is the 11-dimensional Planck
length.  We note that
\footnote{We thank the referee for this comment.}
the spectrum of KK reduction of the eleven-dimensional supergravity
on $S^4$ contains
a massive three-form gauge fields which obeys an ``odd
dimensional self-duality'' condition, or equivalently, two
massless two-form gauge fields under a Hodge duality
\cite{Samtleben:2005bp}. Therefore,
at least in the
large $N$ limit, AdS/CFT predicts 
the existence of two global Abelian 2-form
currents and corespondingly two Abelian global symmetries in the (2,0)
superconformal field theory. 
It is intriguing
to note that there are indeed two $U(1)$ global symmetries within
the symmeries of the
non-abelian tensor gauge fields in the $U(N)\times U(N)$ construction of
\cite{Chu:2011fd}. Such a gauge symmetry in the (2,0) theory
has been predicted to arise from the
$U(N) \times U(N)$ Kac-Moody symmetry of the
multiple self-dual strings worldsheet on the M5-branes
\cite{Chu:2009ms}.

The existence of these global currents is interesting and one can exploit their
properties to learn something about the (2,0) theory.
To do this,
let us introduce a boundary in the M5-branes system. Our AdS/CFT analysis
as performed in the previous section predicts a holographic boundary current
\eq{J-B} in the (2,0) theory in the presence of an external 3-form flux $H$. 
As the supergravity is
maximally supersymmetric, the constant $1/q^2$ in the supergravity action
is not an independent scale, but is related to the AdS radius
\be
1/q^2 \sim R^4
\ee
up to a dimensionless numerical constant.
Since
the 7-dimensional Newton constant
$G_N = G_N^{(11)}/{\rm Vol} (S^4)$ and
$G_N^{(11)} = 16 \pi^7 l_{11}^9$, we obtain
\be \label{b-6d}
b \sim - (R/l_{11})^9 \sim -N^3
\ee
for the two Abelian 2-form currents  in the (2,0) theory 
that are dual to the two massless KK 2-form
gauge fields in the 7-dimensional bulk supergravity.

We notice that in 4-dimensions,
the coefficient $b$ of the boundary current is given by the beta function of
the theory and it
is proportional to the number of degree of freedom that couple to the
$U(1)$ gauge field. 
Here we expect that $b$ to be proportional to the degrees of freedom that
couple to the 2-form gauge field. Our result \eq{b-6d} suggests that
an order of $N^3$ degree of freedom couples to the external
$B_{\m\n}$ field and
the number of degree of freedom in the (2,0) theory is proportional to
$N^3$ for large $N$. We note that a
scaling dependence of $N^3$ also appear in the
entropy of a system of coincident near extremal black 5-branes solution
\cite{Klebanov:1996un}.
However we emphasis that the associated physical mechanism is different:
here there is no horizon in the geometry and a different observable,
a conserved current, is considered.

\section{Discussion}

In this paper we have derived the existence and the
form of a boundary two-form current in the presence of a background 3-form flux
in a 6-dimensional CFT.
The background 3-form flux also induces a Weyl anomaly in the theory.
We derived these results using holographic principle.
An interesting question is whether and how
one may understand these results in terms of
field theory directly.

In 4-dimensions, the induced boundary current
aroused from the magnetization effect of the renormalized vacuum near the boundary. 
Both the current and the Weyl anomaly came from
the quantum loop effects of matter fields in the presence of an external gauge
field background.
The current \eq{J-B}
and the Weyl-anomaly \eq{anom-6d} 
for 6-dimensions should have
a similar origin. Note that there is no obvious way to
couple a point particle to a tensor gauge field $B_{\m\n}$ in 6-dimensions.
However there is a natural way to couple $B$ to a string.
To see this, let us recall that in 4-dimensions, 
the coupling of matter field to external gauge field $A_\m$
can be  obtained by gauging the global
symmetry of the theory.
For example, the $U(1)$ gauge symmetry of a Dirac fermion field 
\be
\psi(x) \to e^{i \a} \psi(x).
\ee
gives rise to the covariant derivative $D_\m \psi = (\del_\m - i A_\m) \psi$ and
the current $J^\m = \psib \g^\m \psi$. In the same way there is a natural way to
construct a covariant derivative for a tensor gauge potential if it is
represented on
a functional $\Psi(C)$ of string/loop. 
First we define the loop derivative 
  \be
  \del_{\m\n} \Psi:=
  \frac{\del \Psi(C)}{\del \sigma^{\m\n}} := \lim_{\d \s^{\m\n} \to 0}
  \frac{\Psi(C+ \d C) -\Psi(C)}{\d \s^{\m\n}},
  \ee
  where $\d \sigma^{\m\n}$ is the infinitesimal area element caused
  by the infinitesimal change in the loop. The derivative exists whenever the
  limit is well defined. 
  $\del_{\m\n}$
  is antisymmetric in the indices $\m, \n$. In general, an arbitrary
  change in the
  phase of the string functional takes the form
  \be
\Psi(C) \to \Psi(C) \exp (i \int_C \a),
\ee
where $\a =  \a_\m dx^\m$ is  an arbitrary 1-form. It is easy to check that
the derivative defined by
   \be \label{cov}
   \cD_{\m\n} \Psi := (\del_{\m\n}  - i B_{\m\n} ) \Psi 
   \ee
   transforms covariantly if $B_{\m\n}$ transforms as
   \be
B_{\m\n} \to B_{\m\n} + \del_\m \a_\n - \del_\n \a_\m. 
\ee
Using the covariant derivative \eq{cov},
one may consider the string field action
\be
S = \int [D x(\s)] \;  \Big(D_{\m\n} \Phi D^{\m\n} \Phi + 
i \Psib \g^{\m\n} D_{\m\n} \Psi\Big),
\ee
where $\Phi$ is a real string field, $\Psi$ is a Weyl spinor string field
and the integration is over all possible closed loops. The action processes
a global  $U(1)$ symmetry which gives the Noether current $J_{\m\n} =
\Psib \g_{\m\n} \Psi$. Our prediction is that the Weyl anomaly \eq{anom-6d}
would arise from
the effective action of the string field theory coupled to an external
3-form flux background. We leave this problem to future work.

\begin{figure}[t]
\centering
\includegraphics[width=5cm]{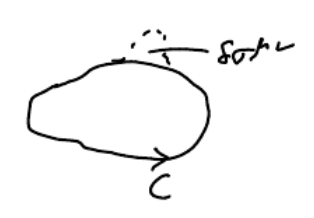}
\caption{Loop deformation}
\label{f2}
\end{figure}

Another interesting question come from the following observation. Recall that 
D-branes in the presence of a
constant 2-form NS-NS $B$-field background is described by a
non-commutative geometry of Moyal type. This can be derived by
considering open string quantisation.
One may expect  a link between the quantum geometry on
the D-brane with the magnetic Casimir effect as both consequence of the
background flux   .
For a M5-brane in the presence
of a constant 3-form $C$-field background, it has been widely speculated that
it must give rises to  some kind of
non-commutative geometry in certain limit.
However much of this speculation is unknown so far. All we
know is that it must reduce to a Moyal type non-commutative geometry
upon a dimensional reduction. 
It may be possible to establish an analogous relation between the
desired quantum geometry and the boundary 2-form current and use the
magnetic Casimir effect to learn about
the quantum geometry.

\vskip7mm
\section*{Acknowledgements}

CSC would like to thank David Berman, Jack Distler, Neil Lambert, Yutaka Masuo
and Douglas Smith for useful comments and questions,
and for the Institute of Mathemaical Sciences, Singapore for hospitality
during the workshop
``String and M-Theory: The New Geometry of the 21st Century''
, where part of this work was carried out.
This work has been 
supported in part by the National Center of Theoretical Science
(NCTS) and the grant 107-2119-M-007-014-MY3 of the Ministry of Science and
Technology of Taiwan.


\appendix

\section{Derivation of Key Relation \eq{key1}}

Consider a BQFT with a well defined effective action.
The integrated
Weyl anomaly $\mathcal{A}$ 
\cite{Duff:1993wm}
\begin{eqnarray}
  \label{A0}
\mathcal{A}=\int_M \sqrt{g} \Big[ g^{\m\n} \la T_{\m\n}\ra-\la g^{\m\n}
  T_{\m\n}\ra\Big].
\end{eqnarray}
can be obtained as the coefficient of the logarithmic
UV divergent term of the
expectation value of the effective action,
\be \label{I0}
I = \cdots + \cA \log (\frac{1}{\e})  + I_{\rm finite},
\ee
where $\cdots$ denotes terms which are UV divergent in powers of the
UV cutoff $1/\e$, and $ I_{\rm finite}$ is the renormalized, UV finite
part of the
effective action.
To derive this result,
let us consider a constant Weyl transformation
$g_{\mu\nu} \to \exp(2 \omega) g_{\mu\nu}$.
Under this transformation, the UV cutoff
transforms as $\e \to \exp(\omega) \e $ and 
the variation of effective action (\ref{I0}) becomes
\be \label{I01}
\delta_{\omega}I = \cdots + \omega (-\cA   +\int_M \sqrt{g}
\la T^{\mu\nu}\ra g_{\mu\nu})+O(\omega^2),
\ee
where we have used $\delta_{\omega} \cA=0$ and
$\delta_{\omega} I_{\rm finite}= \omega \int_M \sqrt{g}
\la T^{\mu\nu}\ra g_{\mu\nu}+O(\omega^2).$
On the other hand, by definition we have 
\be \label{I02}
\delta_{\omega}I = \frac{1}{2}\int_M \sqrt{g}
\hat{T}^{\mu\nu} \delta_{\omega}g_{\mu\nu}
= \omega \int_M \sqrt{g} \hat{T}^{\mu\nu}g_{\mu\nu}+O(\omega^2),
\ee
where $\hat{T}^{\mu\nu}$
is the non-renormalized stress tensor.
We use the hatted  symbol (e.g. $\hat{T}_{\m\n})$
to denote non-renormalized quantity
and un-hatted symbol (e.g. $T_{\m\n}$) to denote renormalized quantity.
Separating  $\hat{T}^{\mu\nu}g_{\mu\nu}$ into the renormalized UV finite part
$\la \hat{T}^{\mu\nu}g_{\mu\nu} \ra$ and divergent part, we have 
\be \label{I03}
\delta_{\omega}I = \cdots+ \omega \int_M \sqrt{g}
\la \hat{T}^{\mu\nu}g_{\mu\nu} \ra+O(\omega^2).
\ee
Comparing the finite part of (\ref{I01}) and (\ref{I03}), we
obtain  the expression (\ref{A0}) for 
$\cal A$ and hence our claim.

Now we are ready to prove the result \eq{key} quoted in the main text of
the paper.
As in \cite{Chu:2018ksb},
let us regulate  the effective action by excluding from its volume
integration a small  strip of  geodesic distance $\epsilon$ from the boundary.
Then there is no explicit boundary
divergences in this form of the effective action, however there are boundary
divergences implicit in the bulk effective action
which is integrated up to distance $\epsilon$.
The variation of effective action with respect to the 2-form potential is
given by
\begin{eqnarray} \label{key1a}
\delta I= \int_{x \ge \epsilon} \sqrt{g} \hat{J}^{\mu\nu}\delta B_{\mu\nu}
\end{eqnarray}
where $\hat{J}^{\mu\nu}=\frac{\delta I}{\sqrt{g}\delta B_{\mu\nu}}$ is the
non-renormalized bulk current.
The renormalized bulk current is defined by the difference
of the  non-renormalized bulk current against a reference one
\cite{Deutsch:1978sc}:
\begin{eqnarray}
  \label{keyJ3}
J^{\mu\n}=\hat{J}^{\mu\n}-\hat{J}_0^{\mu\n},
\end{eqnarray}
where $\hat{J}_0^{\mu\n}$ is the
non-renormalized current
defined for the same CFT without boundary. It is
\begin{eqnarray} \label{key2}
\delta I_{0}= \int_{x \ge \epsilon} \sqrt{g} \hat{J}_0^{\mu\nu}\delta B_{\mu\n},
\end{eqnarray}
where $I_0$ is the effective action of  the  CFT 
with the boundary removed, hence the integration
over the region $x\ge \epsilon$.
Subtract (\ref{key2}) from (\ref{key1a}) and focus on only the
logarithmically divergent terms, we obtain our key formula 
\begin{eqnarray} \label{key-1}
  (\delta \mathcal{A})_{\partial M} = \left(
  \int_{x \ge \epsilon} \sqrt{g}J^{\mu \n}\delta
B_{\mu\n} \right)_{\log (1/\epsilon)},
\end{eqnarray}
where $(\delta \mathcal{A})_{\partial M}$ is the boundary terms in the variations
of Weyl anomaly and $J^{\mu}$ is the renormalized bulk current.
In the above derivations, we have used the fact that $\delta I$
and $\delta I_0$ have
the same bulk variation of Weyl anomaly so that 
\begin{eqnarray} \label{key3}
(\delta \mathcal{A})_{\partial M}= (\delta I-\d I_0)_{\log (1/\epsilon)}.
\end{eqnarray}

\section{Holographic Weyl anomaly}

In this appendix, we investigate the holographic Weyl anomaly for 6d
CFT. Since we are interested in the bulk Weyl anomaly (\ref{anom-6d})
which is irrelevant to the boundary, we focus on the case without
boundary. For simplicity, we focus on the flat space. Then all the
curvatures vanish and only the field strength $H_{ijk}$ contribute to
Weyl anomaly.  According to \cite{Henningson:1998gx}, holographic Weyl
anomaly can be obtained as the UV logarithmic divergent terms of the
gravitational action (\ref{action}).
In the FG gauge, we have
\be \label{metricapp}
ds^2=G_{MN}dx^{M}dx^{N}=\frac{dz^2+\hat{g}_{\m\n}dx^\m dx^\n}{z^2},
\ee
where $\hat{g}_{\m\n}=g_{\m\n}+z^2g^{(2)}_{\m\n}+z^4g^{(4)}_{\m\n}+ \cdots$.
Since we focus on flat space $g_{\m\n}=\eta_{\m\n}$, we have $g^{(2)}_{\m\n}=0$
\cite{Henningson:1998gx,Imbimbo:1999bj}. According to
\cite{Miao:2013nfa}, $g^{(4)}_{\m\n}$ and higher order terms in the
expansions of $g_{\m\n}$ do not contribute to holographic Weyl anomaly
for 6d CFT. Thus, we can set $\hat{g}_{\m\n}=\eta_{\m\n}$ in the following
derivations. Similar to FG gauge (\ref{metricapp}) for the bulk
metric, we take the following gauge for bulk gauge field
\be \label{Bapp}
\cB_{z \m}=0,\ \ \cB_{\m\n}=B_{\m\n}+ z^2 B^{(2)}_{\m\n}+ \cdots
\ee
where $B_{\m\n}$ is the background gauge field for 6d CFT. 

Substituting (\ref{metricapp}),(\ref{Bapp}), together with
$\hat{g}_{\m\n}=\eta_{\m\n}$ into the action (\ref{action1}),
we obtain the logarithmic divergent term as
\begin{eqnarray}  \label{Weylanomalyapp}
I &=&
\frac{1}{16 \pi G_N}
\int dz d^6 x \frac{\sqrt{g}}{z^7} (\cdots - \frac{z^6}{6}
H_{\m\n\l} H_{\a\b\r}g^{\m\a}g^{\n\b}g^{\l\r}),\nonumber\\
&=&
-\frac{1}{96 \pi G_N}
\int_M d^6 x \sqrt{g} H_{\m\n\l}^2 \ln\frac{1}{\epsilon}+ \cdots
\end{eqnarray} 
where $\cdots$ denote
power law divergent terms and regular terms.
From (\ref{Weylanomalyapp}), we can read off the holographic Weyl anomaly 
\be \label{anom-6dapp}
\mathcal{A}=\int_M \sqrt{g}\; \frac{b}{6} H_{\m\n\l}^2
\ee
with $b=-\frac{R}{16\pi G_N q^2}$.
This holographic Weyl anomaly agrees precisely with that obtained in
(\ref{anom-6d}).



 \end{document}